\documentclass[12pt]{iopart}
\usepackage{graphicx}
\newcommand{\vo}{%
          {VO$_2$}}
\newcommand{\vok}{%
          {VO$_2$}}

\newcommand{\bab}{%
           bonding/antibonding }
\newcommand{\agm}{%
          {a_{1g}}}
\newcommand{\egpm}{%
          {e_{g}^\pi}}

\newcommand{\ag}{%
          {$a_{1g}$}}
\newcommand{\egp}{%
          {$e_{g}^\pi$}}

\newcommand{\cc}{%
        c^\dag}    
\newcommand{\ca}{%
        c^{\phantom{\dag}}}  
\newcommand{\svek}{%
        \mathbf}
\newcommand{\vek}[1]{%
        \hbox{\textbf #1}}

\newcommand{\op}[1]{%
        \hbox{\textbf #1}}
\newcommand{\pr}{%
        ^\prime}

\usepackage{iopams}  

\begin{document}

\title[Effective Band Structure of Correlated Materials -- The Case of
\vo]{Effective Band Structure of Correlated Materials
  --\\
  The Case of \vo}

\author{Jan M Tomczak and Silke Biermann}

\address{Centre de Physique Th{\'e}orique, Ecole Polytechnique,
91128 Palaiseau Cedex, France}
\ead{jan.tomczak@polytechnique.edu}

\begin{abstract}
Vanadium dioxide \vo\ and its metal-insulator
transition at T=340K continues to receive considerable interest. The question
whether the physics of the insulating low-temperature phase is dominated by the
Mott or the Peierls scenario, i.e. by correlation or band effects, is
still under debate. 
A recent cluster dynamical mean field theory
calculation\cite{biermann:026404} suggests a combination of both
effects, characterizing the transition as of a correlation assisted
Peierls type. In this paper
we present a detailed analysis of the excitation spectrum of the
insulating M1 phase of \vok, based on this calculation.
We implement a scheme to analytically continue
self-energies from Matsubara to real frequencies, and study
the physics of the strong interactions, as well as
the corresponding
changes with respect to the density functional theory (LDA) band structure.
We find that in the M1 phase life-time
effects are rather negligeable, indeed allowing for an effective
band structure description.
A frequency independent but orbital dependent potential, constructed
as an approximation to the full cluster dynamical mean field
self-energy, turns out to satisfactory reproduce the fully interacting one-particle spectrum,
acting as a scissors operator which pushes the \ag\ bonding and \egp\
bands apart and, thus,  opening the gap.
\end{abstract}



\section{Introduction}
One of the key issues of condensed matter physics, both from the
theoretical and the experimental point of view, is the understanding
of the electronic structure of materials. With the advent of Density
Functional Theory (DFT) within the Local Density Approximation (LDA),
or its gradient extensions, the {\it ab initio} calculation of electronic properties of
astonishingly many classes of compounds became feasible in a reliable
fashion \cite{RevModPhys.61.689}. The manageable complexity of the material ever increases and
even the verge of chemical precision comes into view.

From a conceptual point of view, however, this might appear rather
miraculous, as there are severe, well known, principle limitations.
Indeed DFT is a theory of the ground state only. The interpretation of
the
Kohn-Sham eigenvalues of the effective one-particle system as
excitation energies of the real N-particle system, has no theoretical
justification, and only the striking compatibility with experiment
promotes this interpretation, which is often tacitly assumed.

On the other hand, regarding the original domain of the theory, i.e. ground
state properties, implementations of DFT inevitably have to resort to
approximations, the exchange-correlation potential being unknown. This
leads to the fact that for some even apparently simple
materials, the calculated ground state is found to contradict experimental
findings. This is the case when electronic correlations, i.e. genuine
many-body effects, invalidate the band-picture which originates from an effective
one-particle theory. The most prominent example is the Mott
transition\cite{Imada,bible}, where the correlations prevail upon the itineracy of the
electrons, forcing them to localize. Among the materials where this
mechanism is present are for example Cr-doped $V_2O_3$ or the parent compounds of the high T$_c$ superconductors.
A genuine signature of strong correlations, highly affecting the distribution of spectral weight, is the sensitivity to changes of external parameters, such as temperature or pressure, as 
is indeed observed in experiments such as
transport, photoemission or optics.

The formidable task is to unravel the behaviour of a system in-between
these two limits~: the effective free case (a problem that is diagonal
in momentum space), and the atomic limit (that is diagonal in real
space). In the latter the notion of bands having lost its sense, the 
quantity to discuss is the spectral function. Within a band-theory
it will, for a given momentum, consist of Dirac distributions.
With the increase of correlations these will broaden, due to
scattering effects, inducing finite life-times, and loose weight that will
reappear as incoherent features or satellites at different energies.

In this work, we present for the specific example of the insulating
phase of \vok, a study of how correlations modify the
excitation energies of the system and how, if at all, this can be cast
into an effective one-particle picture. 

To this effect, we implemented a scheme to analytically continue
Matsubara self-energies to the real frequency axis, including for the
first time the continuation of off-diagonal elements. This allows us to
make a detailed analysis of recent cluster dynamical mean field
calculations\cite{biermann:026404}.

The paper is organized as follows~: In section 2 we review the
electronic structure of \vo\ and recent cluster dynamical mean field
calculations. In section 3 we apply the analytical continuation scheme
(that we detail in \ref{anacon}) and discuss the real-frequency
self-energies for the M1 phase of \vok. Section 4 presents a study of
the correlation effects on the excitation spectrum and suggests an
effective one-particle potential which corrects the LDA band
structure.

\section{Electronic Structure of \vok}
\subsection{Peierls versus Mott~: A Reminder}

The phase diagram of vanadium dioxide \vo\ has intrigued condensed matter
physicists for
decades\cite{PhysRevB.11.4383,goodenough_vo2,PhysRevLett.72.3389,sommers_vo2,PhysRevB.10.1801,PhysRevLett.35.873,khomskii_vo2,eyert_vo2,korotin_vo2,tanaka_vo2}.
At high temperature \vo\ is a (bad) metal and the crystal structure is of rutile type. At T=340K the compound undergoes a first-order metal-insulator transition (MIT)\cite{PhysRevLett.3.34} and the structure becomes monoclinic (M1 phase). The paramount characteristic of the low temperature structure is the pairing up of vanadium atoms, which form dimers along the c-axis that tilt out of the axis. Moreover the phase is found to exhibit no local moment behaviour.
There is a long standing debate as to whether the MIT is primarily caused by the structural changes that double the unit cell (Peierls transition), or whether it is correlation effects that drive the system insulating (Mott-Hubbard transition).

\vo\ has a $3d^1$ configuration and hence the key focus lies on the d-orbitals.
The cubic component of the crystal field resulting from the octahedral
coordination of the vanadium atoms in the rutile phase splits the
d-orbitals into an $e_g$ and a $t_{2g}$ manifold, the latter being
lower in energy. The total crystal field further splits the $t_{2g}$
into two \egp\ and a single \ag\ orbital, which is directed along the c-axis. Both types of orbitals overlap, accounting for the metallic character of the rutile phase.
In the M1 phase, where one has four formula units per unit-cell, the
\ag\ orbitals split due to the dimerization into \bab
states. Furthermore, because of an increased hybridization with the
oxygen 2p orbitals owed to the tilting of the dimers, the \egp\ are
pushed up in energy relative to the former centre of the \ag\
orbitals. Both these effects are working in favour of a
gap-opening\cite{goodenough_vo2}.
However, LDA calculations, though decreasing the density of states at
the Fermi level as compared to the rutile phase, fail to capture the
experimental gap (see figure \ref{figZ}), which is only
opened, once non-local correlation effects beyond the LDA are taken into
account\cite{biermann:026404,liebsch:085109,laad:195120}.

\subsection{Dynamical Mean Field Theory in a nutshell}

A theoretical model, that embodies the above mentioned struggle between
localized and itinerant behaviour, or kinetic and potential energy, is
the well-known Hubbard model. In the multi-band case its Hamiltonian reads
\begin{eqnarray}\label{Hub}
\op{H}&=&\sum_{\svek{k},ll\pr,\sigma} H_0^{ll\pr}(\svek{k})\cc_{\svek{k}l\sigma
  } \ca_{\svek{k}l\pr\sigma} +\sum_{{\svek{R},ll\pr,\sigma\sigma\pr}}^{(l,\sigma)\ne(l\pr,\sigma\pr)}
  U_{ll\pr}^{\sigma\sigma\pr} n_{\svek{R}l\sigma}n_{\svek{R}l\pr\sigma\pr}
\end{eqnarray}
where $H_0^{ll\pr}(\svek{k})$ is the one-particle dispersion, l being
an orbital index, and $U$
an on-site Coulomb repulsion that is here limited to density-density
terms only. $\cc_{\svek{k}l\sigma}$, $\ca_{\svek{k}l\sigma}$ are the
usual fermionic creation and annihilation operators and
$n_{\svek{R}l\sigma}$ the corresponding number operators in real space.
In a realistic context, the one-particle dispersion will be taken from
e.g. an LDA calculation. 
 Though the values of the interactions
are often guided by constrained LDA\cite{PhysRevB.44.943}, RPA\cite{PhysRevB.70.195104} or experimental estimates, they
somewhat remain parameters of the technique.

To solve this many-body problem in cases where the dominant
correlation effects are local, a method of choice is the
 Dynamical Mean
Field Theory (DMFT) and its realistic extension (LDA+DMFT)\cite{vollkot}.
The tremendous merit of DMFT is
that it is a non-perturbative technique that is able to accurately describe the
local physics, independent of the interaction strength.
The dynamical aspect allows for an accurate description of spectral
weight transfer to incoherent features.

The basic idea
of DMFT is to replace a lattice problem (or as in the realistic case,
a problem defined within a localized basis set) by an
effective one-site system, coupled to a bath and subject to a
self-consistency condition, analogue to conventional Weiss mean field
theory in statistical mechanics. Contrary to the latter however, the
intervening mean field is energy-dependent, hence the notion of a dynamical MFT. For solving the 
DMFT equations, a variety of techniques are available.
In the case of realistic multi-band calculations, a widely used technique is the  Quantum Monte Carlo (QMC). 
A common feature of QMC implementations is that they work within the finite temperature Matsubara
formalism, i.e. in imaginary time or frequency. Hence, an analytic
continuation to the real frequency axis is needed. 
In \ref{anacon} we describe such a procedure for the DMFT self-energy
in the most general case, that for the first
time includes the continuation of off-diagonal elements. The scheme
thus gives access to the full orbital physics, as well as to non-local
(Cluster DMFT) effects.

\subsection{Recent DMFT Results for \vok}
Recently, both the rutile and the M1 phase of \vo\ were investigated within
DMFT\cite{biermann:026404,laad:195120,liebsch:085109}. Not surprisingly the insulating character of the monoclinic phase proved to be 
not capturable in the standard (single site) LDA+DMFT formalism
without the unphysical creation of local moments. Instead, a
cluster-extension of DMFT (CDMFT), where the solid is replaced not
by a single impurity, but by a two-site cluster of vanadium atoms (surrounded by their
oxygen octahedra) was needed\cite{biermann:026404}. It is of course
very appealing from the physical point of view, to choose the vanadium
dimers as the fundamental unit of the calculation. Herewith
non-local inter-vanadium correlations are treated in a non-perturbative manner.
The calculation in Ref \cite{biermann:026404} used an $N$MTO-Wannier
Hamiltonian\cite{nmto} where all orbitals other than the $t_{2g}$ were
downfolded, an on-site Coulomb repulsion $U=4.0$eV, a Hund's coupling
$J=0.68$eV and $T=770$K.
For solving the DMFT equations, the Hirsch-Fye Quantum Monte Carlo (QMC) algorithm\cite{PhysRevLett.56.2521} was used.
 The insulating nature of the M1 phase is correctly described and the
 resulting local spectral functions of both phases compare favourably
 with
 photoemission and x-ray experiments\cite{PhysRevB.41.4993,PhysRevB.55.4225,kurmaev-1998-10,PhysRevB.20.1546,okazaki:165104,PhysRevB.43.7263,haverkort:196404,eguchi_vo2,koethe:116402}.
The authors concluded that the non-local correlations effectively
assist the Peierls-like transition, the dimerization leading to the
formation of molecular singlets in the \ag\ channel embedded in a bath.

\section{The DMFT Self-Energy on the Real Frequency Axis}
\subsection{Orbital Structure of the Self-Energy}\label{sS}

In the present work, we have implemented an analytic continuation procedure for imaginary frequency DMFT self-energies (see \ref{anacon}), which enabled us to develop a more detailed analysis of the above calculation. With the real-frequency self-energy at hand, interesting quantities become accessible, among which figure momentum-resolved spectral functions and linear response functions, such as the optical conductivity (when neglecting vertex-corrections). 
In the following we shall analyse the correlation induced modifications in the excitation spectrum of M1 \vo\ and discuss whether the result can still be cast into an effective band-picture.

With four formula units in the unit cell, the M1 \vo\ self-energy matrix
acquires the following form~:
\begin{eqnarray}\label{S1}
\Sigma_{unitcell}&=&
\left(\begin{array}{cccc}
\Sigma_{11}&\Sigma_{12}&0&0\\
\Sigma_{21}&\Sigma_{22}&0&0\\
0&0&\Sigma_{11}&\Sigma_{12}\\
0&0&\Sigma_{21}&\Sigma_{22}
\end{array}\right)
\end{eqnarray}
where the blocks $\Sigma_{ij}$ are 3x3 matrices each, and $\Sigma_{ii}$ $(i=1,2)$ are
the local self-energy matrices for a given vanadium atom, and $\Sigma_{12}$ and
$\Sigma_{21}$ are the inter-vanadium, intra-dimer blocks.
Still, when given this additional freedom, the system is found to have
in total only four non-equivalent self-energy elements.
All other elements are zero within numerical precision.
The self-energy for one dimer, i.e. the upper-left or down-right 2x2
block in (\ref{S1}) comes out to be

\begin{eqnarray}\label{S2}
\Sigma_{dimer}&=&
\left(\begin{array}{cccccc}
\Sigma_\agm &0 &0 &\Sigma_{\agm-\agm} &0 &0\\
0 &\Sigma_{\egpm1} &0 &0 &0 &0\\
0 &0 &\Sigma_{\egpm2} &0 &0 &0\\
\Sigma_{\agm-\agm} &0 &0 &\Sigma_{\agm} &0 &0\\
0 &0 &0 &0 &\Sigma_{\egpm1} &0\\
0 &0 &0 &0 &0 &\Sigma_{\egpm2}\\
\end{array}\right)
\end{eqnarray}
i.e. the intra-dimer coupling is entirely due to the \ag\ channel, as can be expected from the discussed symmetry of the system.

\subsection{Interpreting the Self-Energies~: Correlation Effects in \vok}\label{self}

\begin{figure}
\begin{minipage}[b]{0.5\linewidth} 
\centering
\includegraphics[angle=-90,width=1.1\textwidth]{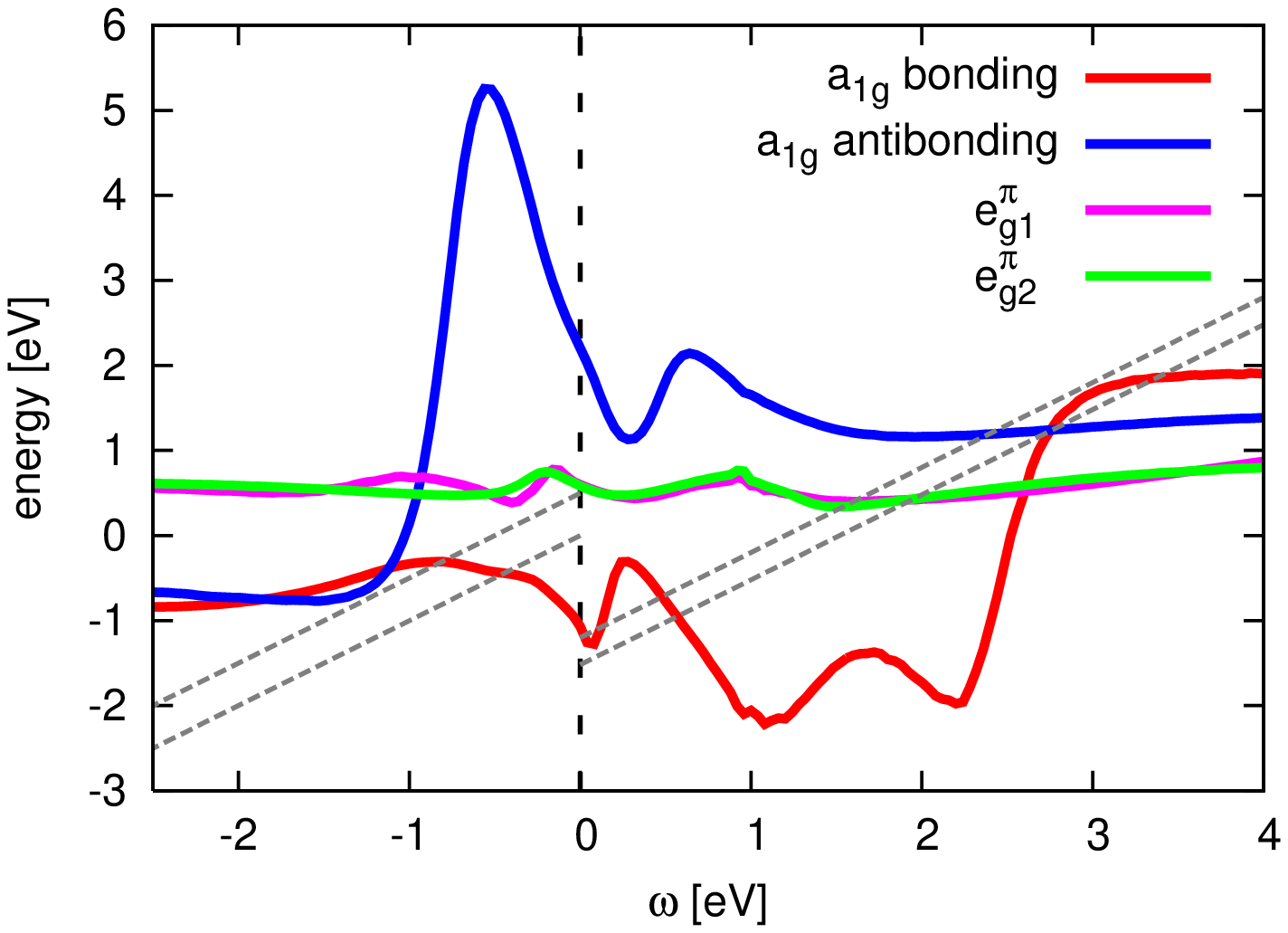}
\caption{Real parts of the self-energy \ag\ \bab basis. The two stripes represent the position and
  bandwidth of the former LDA 
  \ag\ bonding ($\omega<0$), antibonding ($\omega>0$) band. See
  \ref{gd} for a discussion.
}
\label{figS1}

\end{minipage}
\hspace{0.5cm} 
\begin{minipage}[b]{0.5\linewidth}
\centering
\includegraphics[angle=-90,width=1.1\textwidth]{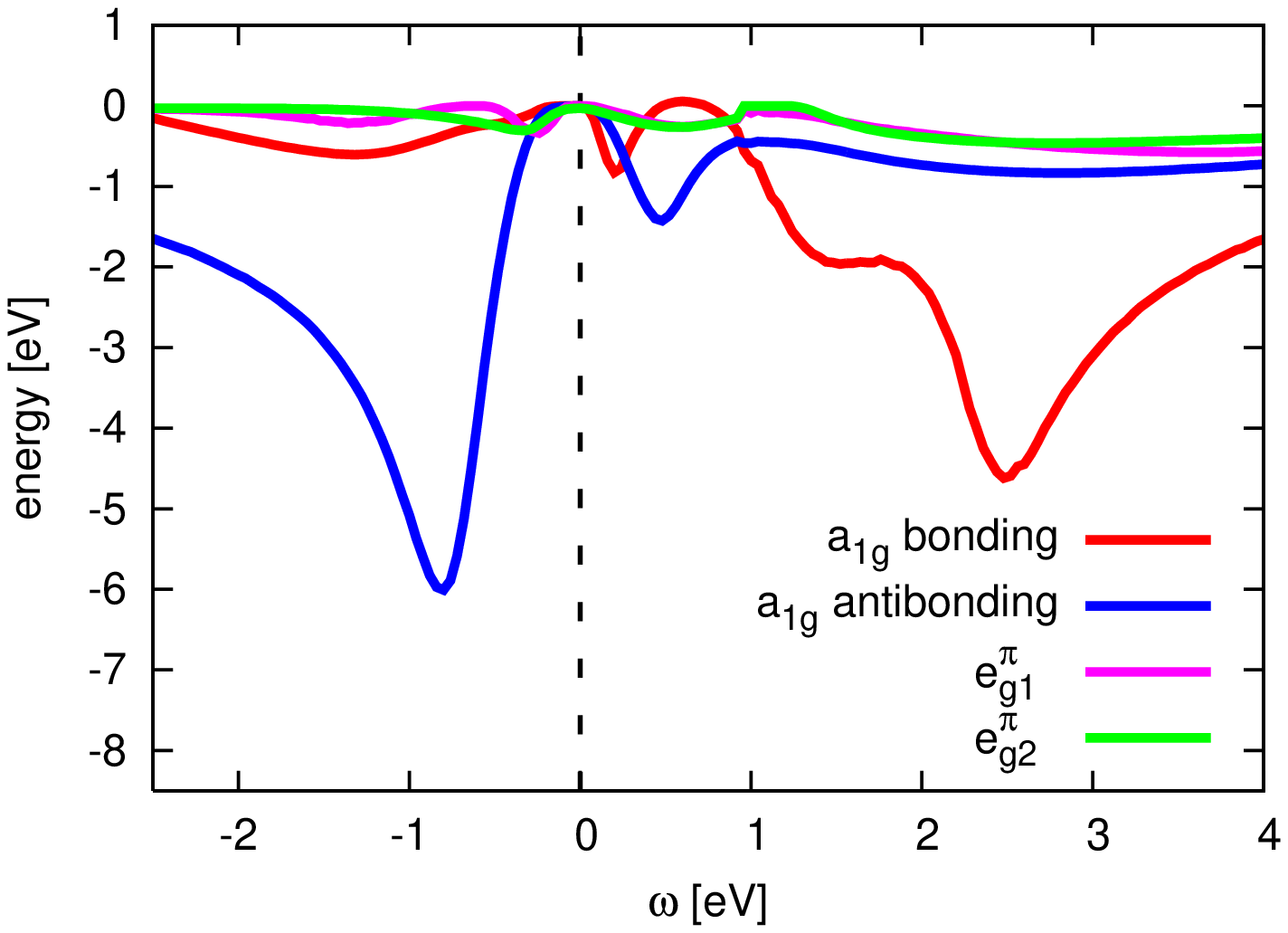}
\caption{Imaginary parts of the CDMFT self-energy of the M1 phase of
  \vo\ in the \ag\ \bab basis. Note that all self-energy elements are regular at zero frequency. See
  \ref{self} for a discussion.}
\label{figS2}
\end{minipage}
\end{figure}

Using the procedure detailed in \ref{anacon}, we compute the
DMFT self-energy on the real frequency axis. Figure \ref{figS1} and \ref{figS2} 
display our result. We have chosen to change the basis of the \ag\
components and to show their \bab combinations
$\Sigma_{b/ab}=\Sigma_{\agm}\pm\Sigma_{\agm-\agm}$, in which the
self-energy is diagonal in orbital space\footnote{We stress that this
  basis transformation can only be done {\it after} the DMFT
  calculation, as in this molecular picture the charges will no longer
  be centered on the atomic sites, and as a consequence the
  Hubbard-Hund Coulomb interaction terms would acquire an untractable form.}.

Several important observations can be made. The suppression of
spectral weight at the Fermi energy can be achieved by two distinct
behaviours of the imaginary parts of the self-energy within the charge
gap: Either they diverge (as is the case of the insulating phase of
the one-band Hubbard model), or they have to vanish within the gap. In
the present case of M1 \vo\ we see, figure \ref{figS2}, that the self-energies are all regular and their imaginary parts vanish within the gap.
Hence it is not a divergence of the effective mass that is responsible for the MIT.

Concerning the \egp\ elements, both the real and the imaginary part, depend only
weakly on frequency. Moreover the imaginary part is globally low in
magnitude, signaling only minor life-time effects.
This is a consequence of the low occupation of these orbitals ($\approx$16\%).
The interesting physics is carried by the \ag\ components. They exhibit a rather selective dependence on frequency:
The $\left[\right.$anti$\left.\right]$bonding component has a quite
flat ($\left[\right.$positive$\left.\right]$ negative) real and an
unimportant imaginary part in the
$\left[\right.$un$\left.\right]$occupied part of the spectrum.
The d$^1$-configuration, with its nearly filled bonding bands, and
empty anti-bonding bands, thus leads again -- but this time in a
non-trivial way -- to only small life-time effects. Despite the
considerable on-site interactions, the complex structure of the
self-energy thus suggests the existence of well-defined one-particle
excitations. As we shall see below, this effective band structure is
however strongly rearranged as compared to the former LDA energies, due to the differences in 
the regions of constant real-parts.

\section{Effective Band Structures~: At the Rescue of the
  Band-Picture}
\subsection{The one-particle Poles}
When neglecting the anti-hermitian parts of the self-energy,
which -- as we have seen above -- are small for the \bab bands, the
excitation energies of the system are given
 by the poles of the one-particle Green's function~:
\begin{eqnarray}\label{Pole}
\det\left(\omega_{\svek{k}}+\mu-H_0(\vek{k})-\Re\Sigma(\omega_{\svek{k}})\right)=0
\end{eqnarray}

In contrast to the DFT equations, the above is not an
eigenvalue problem, due to the frequency dependence of the self-energy.
As a first step we shall discuss a graphical construction for finding solutions of (\ref{Pole}).
\subsubsection{One-particle Excitations:~A Graphical Construction.}\label{gd}
 In the one-band case the problem reduces to
 $\omega_{\svek{k}}+\mu-\epsilon_\svek{k}-\Re\Sigma(\omega_{\svek{k}})=0$.
 Solutions are the intersections of the real-part of the self-energy
 (minus $\mu$) with a frequency stripe of slope one and a width
 corresponding to the dispersion of $\epsilon_\svek{k}$. In the general Hamiltonian case this construction is no longer exact, but might still give qualitative insights. In the present case
we expect this procedure to give a rather good description, when
 working in the \bab basis and limiting ourselves to a discussion of
 the \ag\ orbitals only.
Indeed, if the \egp\ orbitals were empty, this contruction would be exact.

In figure \ref{figS1} we indicate the positions and bandwidths
of the \ag\ bands within LDA. As in the \bab basis the orbital pole
equations approximately decouple, we show the
LDA $\left[\right.$anti$\left.\right]$bonding bandwidth only 
in the region $\left[\omega>0\right]$ $\omega<0$, where relevant
intersections can be found.
Below the Fermi-level, bonding pole solutions appear between -0.5 and -1.0eV, due to intersections with the bonding
part of the self-energy (red) with the stripe representing the
former LDA \ag\ bonding band ($\omega<0$). As a result, the bonding orbital is
considerably pushed downwards in energy.
The \ag\ anti-bonding poles, i.e. the intersection of the (blue) anti-bonding
self-energy and the stripe at $\omega>0$ are pushed beyond 2eV.
Thus, already at this point, we suspect the spectral weight above 2eV
as seen in the local spectral-function of Ref.\cite{biermann:026404}
 to stem from the \ag\ anti-bonding excitation rather than from an upper
 Hubbard band. We will come back to this point in the next section.

\subsubsection{One-particle excitations~: Poles of the Green's Function.}
\begin{figure}[t~!h]
\begin{center}
    \includegraphics[angle=-90,width=0.7\textwidth]{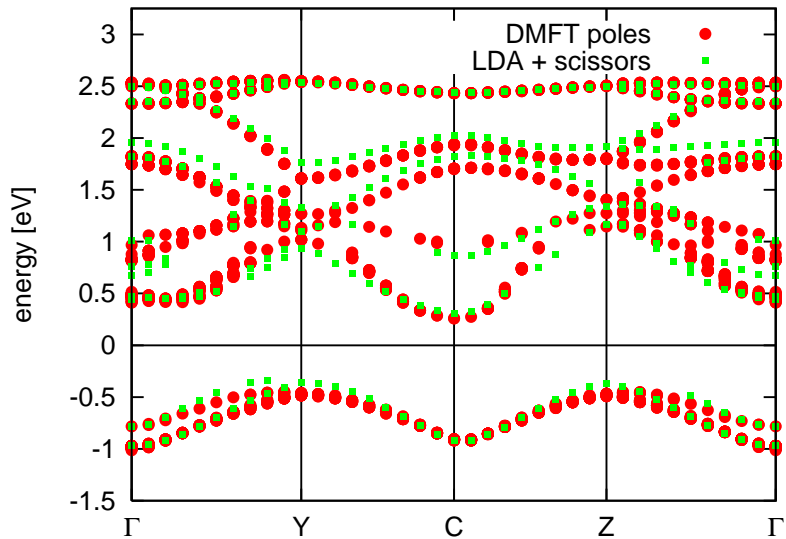}
  \end{center}
\caption{Comparison between the one-particle poles (red) according to
  (\ref{Pole}) and the bands using an optimized orbital dependent
  but frequency-independent effective potential obtained as an
  approximation to the full CDMFT self-energies (green) for various
  symmetry lines. See text for a discussion.}
\label{figPole}
 \end{figure}

To solve the quasi-particle equation (\ref{Pole}), one could in practice scan through frequency and look for sign-changes of the
determinant (\ref{Pole}). However, as along various symmetry-lines, degenerate solutions
exist, we had to introduce a threshold and solutions where
identified as yielding an absolute value of the determinant below this
value. Figure \ref{figPole} shows our results.

On general grounds, there are, at a given $\vek{k}$-vector, at least
as many poles  as there were LDA bands (including possible degeneracies). However, the frequency
dependence of the self-energy may produce additional solutions of
(\ref{Pole})~: In the particle-hole symmetric one-band Hubbard
model, for example, there is one solution in the non-interacting limit, which simply
is the one-particle dispersion, whereas in the atomic limit there are
two poles at $\pm U/2$. In between there is a regime in which three
poles may occur, representing the famous three-peak structure in the
spectral function. In our example of the M1 phase of \vok, we
note that the number of one-particle DMFT poles equals the number of
LDA-bands. In other words, the bands get shifted and reshaped, but
the importance of incoherent features is rather small.
This finding is far from trivial. For example in the metallic rutile phase
of \vok\cite{me_vo2} or V$_2$O$_3$\cite{poter_v2o3}, additional poles
occur, representing incoherent Hubbard bands.

The most important changes as compared to the LDA is the opening of a
gap between the bonding \ag\ and the \egp\ bands. Besides, the
anti-bonding \ag\ band is shifted well beyond the \egp\ ones, as already seen in the graphical discussion above.
In a molecular picture, the \bab splitting of the
\ag-orbitals, due to the dimerization, is tremendously enhanced by the
{\it non-local} intra-dimer correlations that are caused by the {\it local}
interactions.
These results confirm the picture by Koethe \etal who identify a high frequency
XAS intensity peak as the \ag\ anti-bonding band, while a small
low-energy feature in the partial \ag\ spectrum, both seen in the experiment and the CDMFT
calculation results from \egp--\ag\ hybridization \cite{koethe:116402,biermann:026404}.

The neglected life-time effects will in our case broaden the spectra
and shift the positions a little due to their frequency dependence, as
can be seen in the ($\vek{k}$-integrated) spectral
function\cite{biermann:026404}. However they will not, as is e.g. the
case in V$_2$O$_3$\cite{poter_v2o3}, completely suppress spectral
weight for some given pole solutions. Hence, in the current example,
the effective band structure gives a quite accurate guidance to the
physics of the system.

At first sight, it may seem surprising that despite strong local correlations -- the imaginary parts of the
on-site selfenergies reach values of the order of 3eV -- well-defined
one-particle excitations survive. Mathematically spoken, this is the
consequence of a cancellation of local and inter-site self-energies
resulting in nearly negligeable life-time effects for the $\left[\right.$anti-$\left.\right]$bonding
bands at $\left[\right.$positive$\left.\right]$ negative energies, respectively.
Physically, enhanced inter-site hoppings allow the electrons to
avoid the strong on-site repulsion by delocalizing over the dimer, and
the resulting intra-dimer fluctuations reduce the net effects of the interaction\cite{sommers_vo2}.

\subsection{Tailoring of a one-particle Potential beyond the LDA~: Scissors at Work}
As we have seen, the frequency dependence of the \bab self-energy in
 regions where they yield solutions of the pole equation is rather
 unimportant. Hence the question arises as to which extent one could
 derive the effective band structure
from a constant one-particle, albeit orbital-dependent, potential $\Delta$.
To this end we will approximate our dynamical self-energy in a static
 way, by its values at the former LDA band-centres for the \egp,
  and by the values at the pole energies for the \ag\ orbitals.
This means we choose $\Delta_{\egpm_1}=\Re\Sigma_{\egpm_1}(0.5eV)-\mu=0.48eV$,
 $\Delta_{\egpm_2}=\Re\Sigma_{\egpm_2}(0.5eV)-\mu=0.54eV$ for the \egp\
 components, and $\Delta_b=\Re\Sigma_b(-0.75eV)-\mu=-0.32eV$, $\Delta_{ab}=\Re\Sigma_{ab}(2.5eV)-\mu=1.2eV$
 for the \ag\ \bab components respectively.
As a consequence, there will be a downshift of the \ag\ bonding band
 ($\Delta_b<0$), which, together with an upshift of the \egp\ bands
 ($\Delta_{\egpm_{1/2}}>0$) is responsible for the opening of the gap.
Thus, these considerations put numbers to Goodenough's
 picture\cite{goodenough_vo2}.
However, the effect caused by the re-hybridisations with the
 oxygen bands, as described in the introduction, is supplemented by
 the following~: The correlations cause a depopulation of the \egp\
 bands, in favour of the \ag\ bonding one, as thereby the electrons are
 effectively avoiding the on-site Coulomb interaction.
 Moreover,
 the anti-bonding \ag\ band is pushed up in energy way beyond all the \egp,
 tremendously enhancing the \bab splitting.

After transforming back into the original basis, we solve the eigenvalue
problem for $H_0(\vek{k})+\Delta$. The results are shown in
figure \ref{figPole} along with the exact poles from above.
The shiftings of the bands, owed to the additional potential, as
 described above, are, besides minor
details, in excellent agreement with the CDMFT poles.

The above one-particle potential acts like a scissors-operator and
is somewhat reminiscent of results obtained within the GW
approximation\cite{hedin}. Indeed, within a model GW approach, a gap
was found to open in the M1 phase\cite{PhysRevB.60.15699}. Furthermore, preliminary results of a
fully {\it ab initio} GW calculation suggest that the GW
approximation is as a matter of fact suited to describe the insulating
character in the present case.

\subsection{Linearizing the Self-Energy}
Though disposing of the fully frequency dependent DMFT self-energy, we find it instructive to compare with an approximate low-frequency expansion, one would employ if one had used a many-body technique that gives only information about the
renormalization of low lying coherent excitations, such as e.g. 
slave-boson theories.
 The renormalized
band structure is then given by the solutions $\omega_\svek{k}$ of
\begin{eqnarray}\label{Z}
\det\left(\omega_{\svek{k}}-Z\left[H_0(\vek{k})+\Re\Sigma(i\omega\rightarrow0)-\mu\right]\right)=0
\end{eqnarray}
 where
$Z^{-1}=
1-\partial_\omega\Im\Sigma(i\omega)|_{i\omega\rightarrow0}$, and
$\left[i\Im\Sigma\right]$ $\Re\Sigma$ is the
$\left[\right.$anti-$\left.\right]$ hermitian part of the self-energy on the imaginary axis.
This approach, exact at the Fermi-level and normally identified with the Fermi-liquid regime, may
in principle be
applicable in our case to the insulating phase of \vo\ as the
self-energies do not diverge at low frequency, as is the case in the
Mott-insulating phase of the one-band Hubbard model.
Albeit, the quantity Z should, in this context, not be identified with
the usual quasi-particle weight.

\begin{figure}[t~!h]
\begin{center}
    \includegraphics[angle=-90,width=0.7\textwidth]{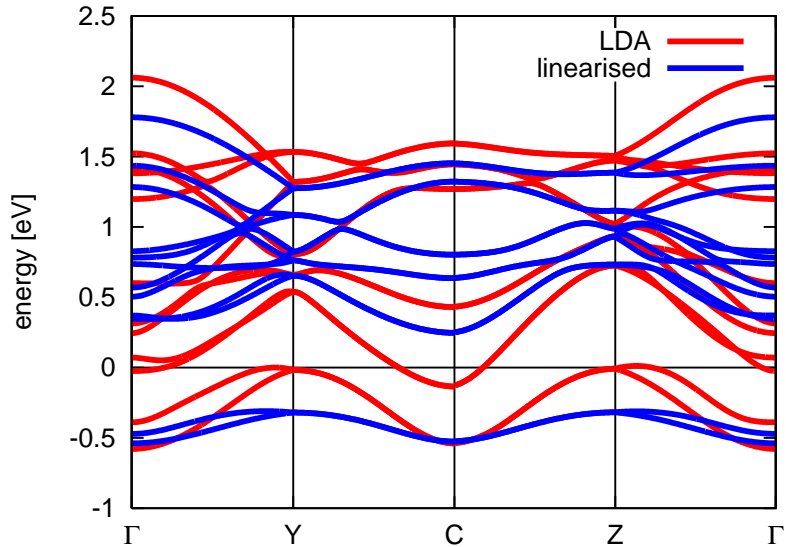}
  \end{center}
\caption{Comparison of the LDA-bands (red) with the bands resulting from the
linearization scheme of (\ref{Z}) (blue)}
\label{figZ}
 \end{figure}

As can be seen in figure \ref{figZ}, a charge gap is found to open,
thus correctly describing the insulating behaviour; however
when comparing to the one-particle poles, we remark huge discrepancies:~
The overall bandwidth is too small, the bands too flat and the \bab splitting
lacks the tremendous increase found before.
This is natural, since foremost the involved slopes of the self-energy
at low-frequency overestimate the renormalizations towards the
Fermi-level, whereas at high enough energy the self-energies are
essentially constant.

\subsection{Conclusions}
In conclusion, we have presented a detailed analysis of the excitation
spectrum of the M1 phase of \vok, with special emphasis on an effective
band structure description. Indeed, we found life-time effects to be
rather negligeable and the correlation effects to only shift and
reshape the Kohn-Sham energies with respect to the starting LDA calculation.
Furthermore, by approximating the dynamical self-energy, we were able to construct a frequency-independent, though
orbital-dependent, one-particle potential, that, when added to the LDA
reproduced the DMFT excitation energies to a quite satisfactory
extent. We remark that due to the orbital-dependence of this
potential, it cannot be viewed as an improved density functional.

Hence, despite the undeniable presence of strong local correlations, the
system retains the coherence of its excitations by means of
intra-dimer fluctuations, to an extent that the physics of the
compound is indeed dominated by the Goodenough-Peierls-picture.
The role of correlations consists in (i) pushing the \ag\ anti-bonding
band beyond the top of the \egp, consistent with the experimental
findings, and, more importantly, in (ii) enhancing
the \ag\ bonding -- \egp\ splitting due to an effectively reduced
Coulomb repulsion in the \ag\ bonding band, hence favouring the depopulation
of the \egp\ bands. The latter results in the opening of the gap.
Thus, as a matter of consequence it {\it is} the correlations that are
responsible for the insulating state, albeit they cause it in a
rather specific fashion.

\ack
We thank A. I. Poteryaev, A. Georges and A. I. Lichtenstein for useful
discussions and for collaborations which resulted in \cite{biermann:026404},
which was the starting point of the present work. Further, we
appreciated discussions with F. Aryasetiawan, M. Gatti, L. Reining,
G. A. Sawatzky, and A. Tanaka. JMT kindly acknowledges support by the
Japan Society for the Promotion of Science (JSPS) under grant PE05050.  
This work was supported by Idris, Orsay, under project No. 061393.

\appendix
\section{Analytical Continuation of the DMFT-Self-Energy}
\label{anacon}

Whereas for the continuation of the Green's function from the Matsubara
formalism to real frequency a numerical technique, called the Maximum Entropy algorithm
(MaxEnt)\cite{maxent} has become a standard approach, the
analytic continuation of the self-energy has only recently been envisioned in
the context of realistic (i.e. multi-band) calculations\cite{biermann_rep,bluemer,anisimov:125119,nekrasov:155112}.

In this work we have implemented a scheme that is sufficiently general
to work in the multiorbital cluster DMFT case, and which, in particular, allows for 
off-diagonal self-energy elements. The following sketch outlines our continuation procedure~:

\begin{center}
\begin{tabular}{cc}
QMC~: $G_{ll\pr}(\tau)$ \\
$\downarrow$  & MaxEnt\\
$A_{ll\pr}(\omega)=-1/\pi\Im G_{ll\pr}(\omega)$\\
$\downarrow$  & Kramers-Kronig\\
\end{tabular}
\end{center}
\begin{equation}\label{root}
G_{ll\pr}(\omega)=\sum_k [w+\mu-H_0(\vek{k})-\Sigma(\omega+i0^+)]^{-1}{}_{ll\pr}
\end{equation}
\begin{center}
\begin{tabular}{cc}
\phantom{ddddddd}$\downarrow$ \phantom{ddddddddd} & Root-finding\\
$\Sigma_{ll\pr}(\omega)$
\end{tabular}
\end{center}

Starting from the local Green's function in imaginary time, we perform the aforementioned MaxEnt algorithm, yielding the spectral function, which is proportional to the imaginary part of the real-frequency Green's function. The real part of this quantity is related to the former by Kramers-Kronig transformation. 
The by far most difficult step in this scheme is the last one. Whereas in the one-band
Hubbard model on the Bethe-lattice an exact expression for the self-energy in
terms of the local Green's function exists ($\Sigma=\omega+\mu-t^2G-1/G$), in the general multi-band case a
 multi-dimensional root-finding procedure\cite{PhysRevB.51.11704} has to be employed to solve
(\ref{root}). In the present case, of four formula units in the
cell, the Hamiltonian is a 12x12 matrix and the self-energy has the
structure shown in section \ref{sS}.

In the degenerate case, the k-sum in (\ref{root}) can
be replaced by an integral over the density of states and one ends up with
the simpler task of inverting the Hilbert transform.

In the most general case, the starting Green's function will furthermore have
off-diagonal
elements in orbital space. As the standard implementation of the
MaxEnt-algorithm in this context has, to our knowledge,
so far been used for diagonal elements only, some comments are in
order.
The equation that is to be inverted reads
\begin{equation}\label{maxent}
G_{ll\pr}(\tau)=\int d\omega A_{ll\pr}(\omega)K_\beta(\tau,\omega)
\end{equation}
with $K_\beta(\tau,\omega)$ being the usual MaxEnt kernel\cite{maxent}.

In contrast to the diagonal elements of the spectral function, which are
normalized to one and are positive at all frequencies, off-diagonal
elements have vanishing zeroth moment and hence change sign as a function of $\omega$. 

For mending this difference, the idea is simply to add a function f($\omega$)
of zeroth moment one to $const\cdot A(\omega)$, and calculate the corresponding shift in
G($\tau$).
 However this function 
 has to be chosen such, that the resulting fictitious
spectral function is non-negative for all frequencies.
To this end, we construct it in the following way.
Given the orthogonal orbital transformation
$\widetilde{A}=U^\dag AU$ with

\begin{equation}
\begin{array}{c}
\\
U_{ll\pr}=\frac{1}{\sqrt{2}}
\end{array}
\begin{array}{c}
\phantom{\downarrow}\downarrow l\pr\\
\left( \begin{array}{ccccc}
1 &  & -1 &  &  \\
 & \ddots &  &  &  \\
1 &  & 1 &  & \\
 &  &  &\ddots & \\
 &  &  &  & 1\\
\end{array}
\right)
\end{array}
\quad
\begin{array}{c}
\\
\leftarrow l\\
\end{array}
\end{equation}
the ll-element ($l{{\tt<}\atop{\tt>}}l\pr$) of the spectral function in the new basis reads
\begin{equation}\label{A3}
\widetilde{A}_{ll}=\frac{A_{ll}+A_{l\pr l\pr}}{2}\pm A_{ll\pr}
\end{equation}
and is thus per construction a proper diagonal element of the spectral
function in the rotated basis, i.e. in particular it is positive for all
$\omega$. Therefore $f=1/2 (A_{l\pr l\pr}+A_{ll})$ is a suitable shift 
for $\pm A_{ll\pr}$ in the original basis. 

Hence, after having continued the original diagonal elements, one
constructs as input the function
\begin{equation}
\widetilde{G}_{ll}(\tau)=\left(U^\dag G(\tau)U\right)_{ll}=\int d\omega
\widetilde{A}_{ll}(\omega)K_\beta(\tau,\omega)
\end{equation}
from which, using (\ref{A3}) the off-diagonal spectral function is then deduced.
We note that in the present case the above rotation has the same
geometrical interpretation as the transformation into a \bab basis,
as used for the \ag\ self-energies of \vok, but the construction is more general and does not require special symmetry properties of the system.

\section*{References}

\end{document}